\documentclass[9pt,twocolumn,twoside,lineno]{revtex4-1}

\usepackage[utf8]{inputenc}
\usepackage[T1]{fontenc}

\usepackage{natbib}
\usepackage{amsmath,graphicx,color,enumitem,hyperref}
\usepackage[FIGTOPCAP,raggedright,nooneline,bf,footnotesize]{subfigure}
\usepackage[version=4]{mhchem}
\newcommand{\figref}[1]{Fig.~\ref{#1}}

\begin{document}

	\title{A single photon detection system for visible and infrared spectrum range}

	\author{Alexander Divochiy}
	\email[]{These authors contributed equally to this work.}
	\affiliation{LLC   SCONTEL,   5/1-14  L'va  Tolstogo  st,  Moscow  119021,  Russian
		Federation}
	\affiliation{Moscow State  University of Education (MSPU), 1/1 M. Pirogovskaya st., Moscow 119991, Russian Federation}
	
	\author{Yury Vakhtomin}
	\author{Pavel Morozov}
	\affiliation{LLC   SCONTEL,   5/1-14  L'va  Tolstogo  st,  Moscow  119021,  Russian
		Federation}
	\affiliation{Moscow State  University of Education (MSPU), 1/1 M. Pirogovskaya st., Moscow 119991, Russian Federation}
	\author{Konstantin Smirnov}
	\author{Philipp Zolotov}
	\affiliation{LLC   SCONTEL,   5/1-14  L'va  Tolstogo  st,  Moscow  119021,  Russian
		Federation}
	\affiliation{Moscow State  University of Education (MSPU), 1/1 M. Pirogovskaya st., Moscow 119991, Russian Federation}
	\affiliation{ National Research University Higher School of Economics, Moscow 101000, Russian Federation}
	
	\author{Marta Misiaszek}
	\email[]{These authors contributed equally to this work.}
	\affiliation{Faculty of Physics, Astronomy and Informatics, Nicolaus Copernicus University, Grudziadzka 5, 87-100 Toru\'{n}, Poland} 
	
	\author{Piotr Kolenderski}
	\email{kolenderski@fizyka.umk.pl}
	\affiliation{Faculty of Physics, Astronomy and Informatics, Nicolaus Copernicus University, Grudziadzka 5, 87-100 Toru\'{n}, Poland} 
	
	
	
	
	\begin{abstract}
		We demonstrate niobium nitride based  superconducting single-photon detectors  sensitive in the spectral range  $452$ nm -- $2300$ nm. The system performance was tested in a real-life experiment with correlated photons generated by means of spontaneous parametric down conversion, where one of photon was in the visible range and the other was in the infrared range.  We  measured a signal to noise ratio as high as $4\times 10^4$ in our detection setting. A photon detection efficiency as high as $64$\% at $1550$ nm and $15$ \% at $2300$ nm was observed.
	\end{abstract}

	\maketitle
	
	\section{Introduction}
	
	Recent advances in field of the superconducting single-photon detectors (SSPD) have introduced many conceptual ideas which allow us to progress their capabilities. Specifically, quantum efficiency (QE) of SSPDs and system detection efficiency (SDE) of SSPD-based systems have been increased to near-unity values at telecom wavelength range\cite{marsili_93,MoSi_HQE,Zv_HQE,zhang_HQE,Smirnov2018}. The timing resolution can reach values below $17.8$ ps \cite{Shcheslavskiy2016}, which is inaccessible for other single-photon detection technologies, such as silicon and InGaAs/InP photodiodes.  In addition to that, SSPDs are considered to be the only available type of the single-photon detectors capable of operation in the wide spectral range, starting from near-UV \cite {SSPD_uv}up to mid-IR \cite{Tarh_5,marsili_05-5,Chen2017,Marsili_13} wavelengths. On the other hand, the small size of SSPDs makes it challenging to couple photons to its sensitive area. Typically, it requires single-mode fibers mounted directly to SSPD. 
	
	As it is known, SDE of SSPDs  depends on two quantities, the absorption and the intrinsic quantum efficiency. The absorption depends on the properties of optical structures, incident field and photon coupling with sensitive area of a detector. The intrinsic quantum efficiency is meant as the probability of resistive state formation due to an absorption event. It depends on the characteristics of the superconducting film. Both the parameters should be maximized in order to make SDE as high as possible. The former can be maximized by proper optical cavity fabrication and can reach near-unity values at particular narrow wavelength range. On the other hand, the optimization of the intrinsic quantum efficiency is a more complexed task. It is higher at lower working temperature \cite{Tarh_5,Marsili_13} and for narrow superconducting stripes \cite{marsili_05-5}, as for the materials with lower superconductor energy gap of superconducting material. However, the limits of the gap are dictated by the device operating temperature. 
	
	The very first implementation of SSPD with single-mode fluoride fiber input showed SDE below $1$\% at $ 2200$ nm \cite{Beveratos}. The detectors were cooled in liquid \ce{He_{4}} down to $1.7$K. More recently the use of multi-mode chalcogenide fiber (IRF-S-100) for transmission of MID-IR photons allowed  detection of photons in the range  up to $7\ \mu$m \cite{Chen2017}. However, due to the mismatch in size between the fiber core and the detector only $ ~1$\%  of the  light incident  on the fiber input reached the SSPD. As a material for fabrication of superconducting structure \ce{WSi} was used. It has a lower critical temperature as compared to  niobium nitride (NbN)  so it can potentially be more useful for mid-IR applications. This is the price of having a working temperature in sub-Kelvin range. 
	
	In our work, we show that the single photon detecting system with NbN-based SSPDs, single-mode fibers and Gifford-McMahon cryocooler (Sumitomo RDK-101D), reaches characteristics which are superior to all previously published such systems for the wavelength range up to $2300$ nm.  In order to increase the intrinsic  quantum efficiency of the detectors, we used an approach similar to the one presented earlier \cite{Smirnov_HQE}, which allows us to fabricate disordered NbN films. In our previous work \cite{Zolotov_mir} we presented preliminary tests of detectors optimized for the $1600-2200$ nm range.
	In addition, it was calculated that in the mid-IR wavelength region the intrinsic quantum efficiency of the device could be as high as $10$\% . Presented coupling schemes could not find their place in the routine experiments because they imply significant prevalence of black body radiation over the signal of interest. To investigate exact values of the SDE of SSPD in the spectrum range up to $2300$ nm we suggest two main novelties with respect to the detector systems used in the telecom range.  An optimized solution for the coupling scheme requires the use of SMF2000 single-mode fiber with the NbN devices.  This significantly increases the allowed operation temperature of the system, which is possible to reach in a typical  close-cycle  cryostat.
	
	\section{SSPD characterization}
	We used our standard technique \cite{Smirnov_HQE,Zolotov_mir} for fabrication of the devices analysed here. The aim was to get higher absorption above $1550$ nm which was done by thickening the dielectric \ce{Si_3N_4} layer. This layer has a refractive index of $2.3$ and a thickness of $185$ nm. We begin with the analysis of the absorption of our devices as a function of incident photon wavelength.
	
\begin{figure}[t]
		\centering
		\begin{tabular}{cc}
			\subfigure[$800$ - $1700$ nm]{\includegraphics[width=0.5\columnwidth]{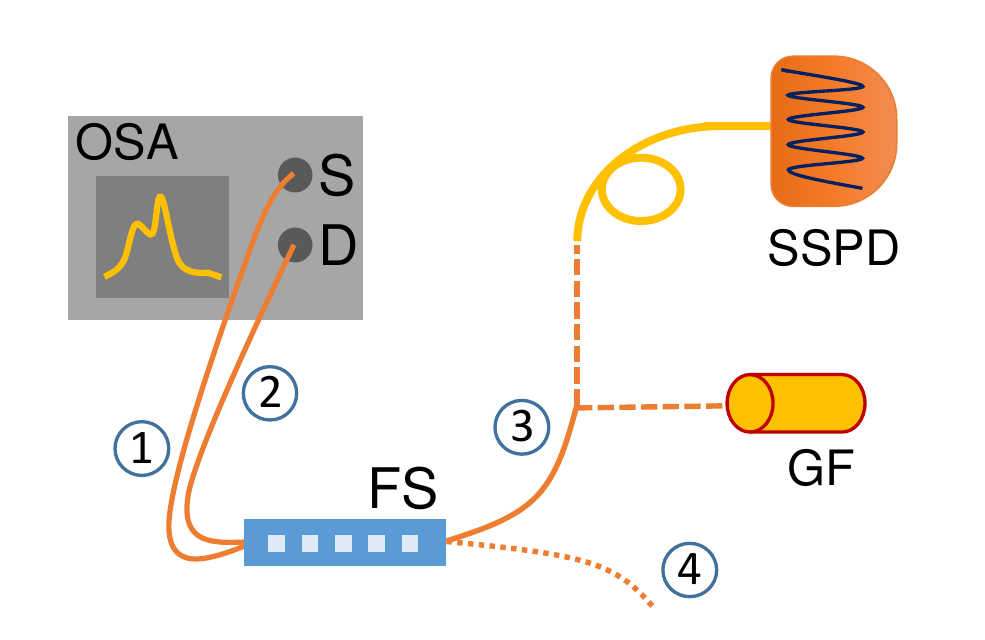}} &
			\subfigure[$1300$ - $2300$ nm]{\includegraphics[width=0.5\columnwidth]{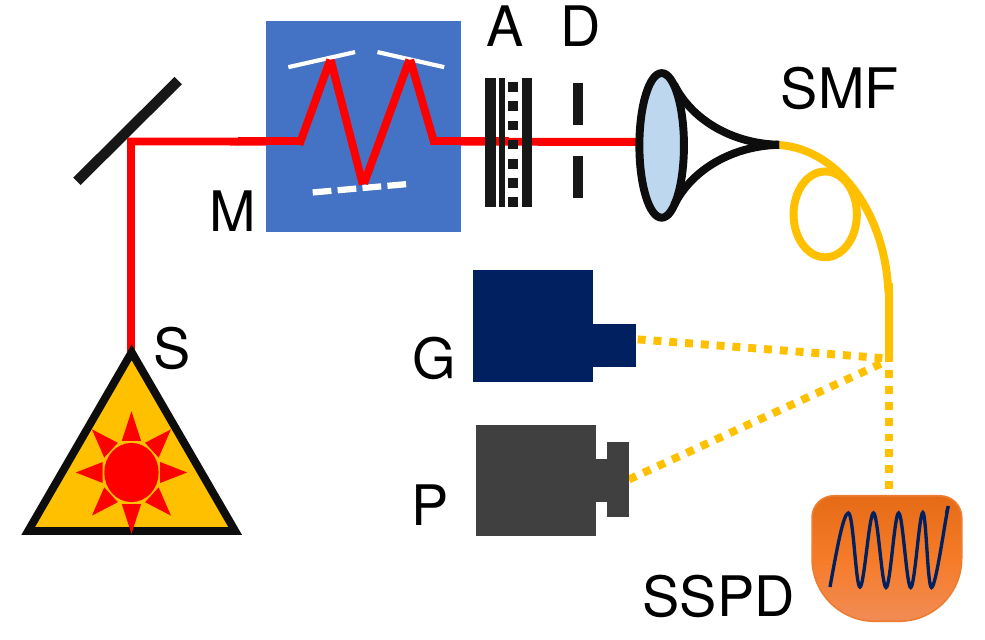}}
		\end{tabular}
		
		\caption[]{Absorption measurement setups. For the range of (a) $800-1700 $ nm: OSA -- Optical Spectrum Analyzer (Keysight HP 70950A), S -- radiation source, D -- detector, FS -- 50:50 fiber splitter for $1550 $ nm, GF -- gold-tip ferule, SSPD  (b) $1300-2300 $ nm: S -- photon source (xenon lamp), M -- monochromator, A -- attenuator, D -- diaphragm, detectors: G -- Golay cell, P -- power meter (Ophir PD-300 IRG), SSPD. }
		\label{fig:setup:absorption}
	\end{figure}

	We investigate the absorption for the spectral range of $800-2300$ nm. It has been done in two steps by using two measurement apparatus depicted in \figref{fig:setup:absorption}:   (a) first for the range of $800-1700$ nm and (b) the other one for the range of $1300-2300$ nm. 
	
	For the analysis of the spectral absorption of our structures we used an optical spectrum analyzer (OSA)  with built-in broadband radiation source allowing for measurement in the spectral rage of $800-1700$ nm, see \figref{fig:setup:absorption}(a). We used a fiber splitter (50:50 distribution at $ 1550$  nm) for simultaneous illumination and detection of the absorption and reference spectrum.  The broadband light source was connected to port number $1$ and OSA monitored the output port $2$.  Port number $3$ was connected to either gold mirror or SSPD under the study. The reflected light  was sent back and registered by OSA. Port number $4$ of the splitter was not used. The fiber splitter's partition coefficient is wavelength-dependent and therefore  in order to perform absorption measurements of superconducting structures, the reference spectra with a gold mirror was acquired. We deposited  $100$ nm thick gold layer on the standard ferule to make sure that there is no air gap between the gold mirror and the fiber port $3$. Thanks to that, in the scheme with SSPD detector, we could calculate the absolute values of the absorption.
	
\begin{figure}[h]
		\centering
		\includegraphics[width=0.9\columnwidth]{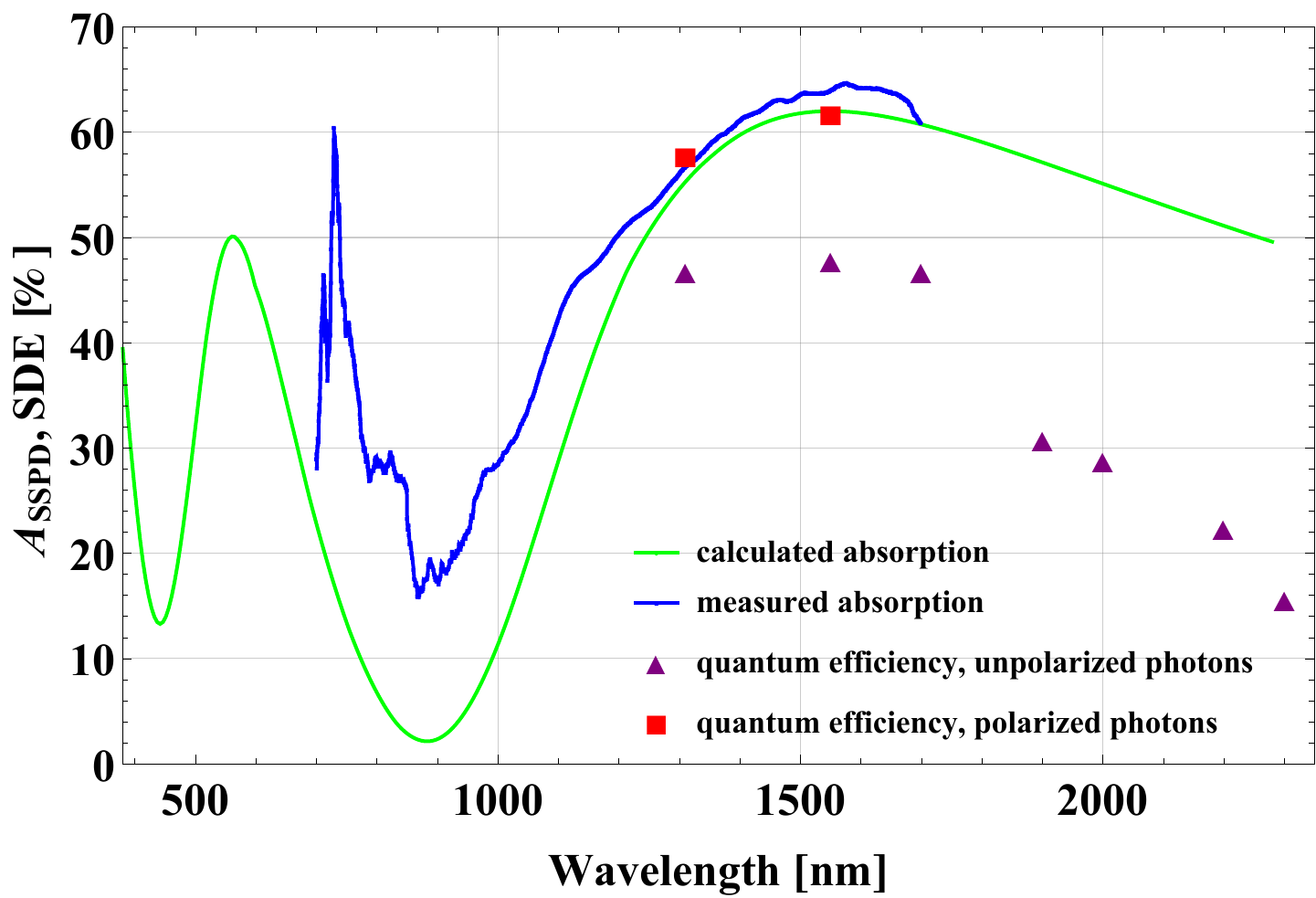}
		\caption[]{The spectral absorption, $A_\text{SSPD}$, and system detection efficiency, SDE.}
		\label{fig:AbsQE}
	\end{figure}

	The measured spectra $P=P(\lambda)$ on port $2$ of the fiber splitter, FS, see \figref{fig:setup:absorption}, in the case of SSPD and gold-tip ferule, GF, are $P_\text{SSPD}(\lambda)=T^2_\text{BS}R_\text{SSPD}\cdot P_0(\lambda)$ and $P_\text{GF}(\lambda)=T^2_\text{BS}R_\text{GF}\cdot P(\lambda)$, respectively.  Here $P(\lambda)$ is the light  source spectrum, $T_\text{BS}$ is the transmission coefficient of the fiber splitter and $R_\text{SSPD}$, $R_\text{GF}$ are the reflection coefficients of the superconducting structures and gold-tip ferule.  Assuming the perfect reflection from the gold mirror (GF), we can put $R_\text{GF}=1$. As a result, we get the relation ${P_\text{SSPD}}/{P_\text{GF}}=R_\text{SSPD}$. By definition, the absorption coefficient reads $A_\text{SSPD}=1-R_\text{SSPD}$, so finally one gets $A_\text{SSPD} = 1-{P_\text{SSPD}}/{P_\text{GF}}$. The measured absorption is depicted in \figref{fig:AbsQE} with a blue curve.
	
	In order to investigate SDE above $1700$ nm,  another measurement technique was developed. It is presented in \figref{fig:setup:absorption} (b). The experimental setup was based on an IR spectrophotometer (IKS-19). Light from the xenon lamp was led to the single diffraction grating with  $300$ grooves/mm. The output slit of $1$ mm width defines \mbox{$~5$ nm} bandwidth of the output radiation. Next, the radiation is attenuated by a set of 1-inch Si wafers with deposited thin NbN films of thicknesses $5$, $10$, $20$ and $30 $ nm and transmittance of $ ~0.4$, $~0.1$, $~0.03$ and $ ~0.01$, respectively. After the attenuator, we used a $0.5$ mm diaphragm and the SM2000 single-mode fiber to collect free-space radiation. 
	
	Due to the fact that transmittance of attenuators depends on wavelength, we calibrated each of them on each wavelength of interest by measuring power with and without attenuation. In order to measure power, we used Golay cell (Tydex, GS-1D \footnote{\url{http://www.tydexoptics.com/}}), which has $6$ mm diameter input covered by diamond window. Furthermore, it has noise equivalent power NEP$ = 10^{-9} W\cdot{Hz}^{-\frac{1}{2}}$ , which does not vary in spectral range of interest. As a result, it perfectly meets the requirements of our experiment. In order to use Golay cell as a power meter we measured its responsivity, which was estimated to be $6800$ V/W. The responsivity measurement was performed using $1550$ nm diode laser and a power meter (Ophir, PD-300 IRG). In order to define the part of the optical power, which can be collected to the single-mode fiber, we measured power at the fiber output directly with the power meter in the range of $1310-1600$ nm. The power meter accuracy is $~1 $ pW in this spectral range. By comparing these results with those achieved on the output of the monochromator with the Golay cell we found that the coupling coefficient is equal to $0.0189(8)$. We assume this coefficient to be constant for the spectral range of $1310-2300$ nm.

	For the measurements, the target power of $0.5-2$ pW was adjusted using a set of attenuators. Single mode fiber, SM2000, was connected to the optical input of the SSPD. The SDE was calculated by measuring the number of electrical pulses using an universal frequency counter (Agilent, A53131). The results of the experiment are presented in \figref{fig:AbsQE} with purple triangles. Note, the difference between the SDE for polarized and unpolarized light stems from the meander-type geometry of the SSPD  \cite{Anant2008, Dorenbos2008}. The system detection efficiency at $1310$ nm and $1550$ nm was approximately $0.45$. Then efficiency decreases with decreasing photon energies.
	In order to estimate the maximal SDE values we also measured quantum efficiency with a linearly polarized light source at available wavelengths. In this case the best values for our detectors were $0.58$ and $0.62$ at $1310$ nm and $1550$ nm, respectively. 

	\subsection{Simulation of absorption in the range of $380-2300 nm$}
	In our work spectral simulations of absorption for the range of $380-2300$ nm were performed with OpenFilters software \cite{OpenFilters}. We adopted parameters of our sample from previous work\cite{Anant2008}. Note that the optical properties of thin \ce{NbN} films vary\cite{Anant2008,Semenov2009a}, which could be explained by differences in stoichiometry of the films. The absorption coefficient was calculated for unpolarized radiation incident on a Au/\ce{Si_{3}N_{4}}/NbN structure ($80$ nm/$185$ nm/$5$ nm thick respectively) from \ce{SiO_{2}} medium. In our simulation we assume that the film is uniform, because the etched regions are much smaller than the wavelengths.  The NbN meander pattern with filling factor of $0.5$ is ignored. The green curve in \figref{fig:AbsQE} presents the simulated absorption spectral dependence for the superconducting structure. A fair agreement between the measured and simulated characteristics in the region of $1300-1700$ nm confirms the accuracy of our back-reflection method, which justifies its application for future spectral data analysis.

	\section{Application for correlated photon pairs}
	\subsection{Experimental setup}
	
	\begin{figure}[h]
		\centering
		\includegraphics[width=0.9\columnwidth]{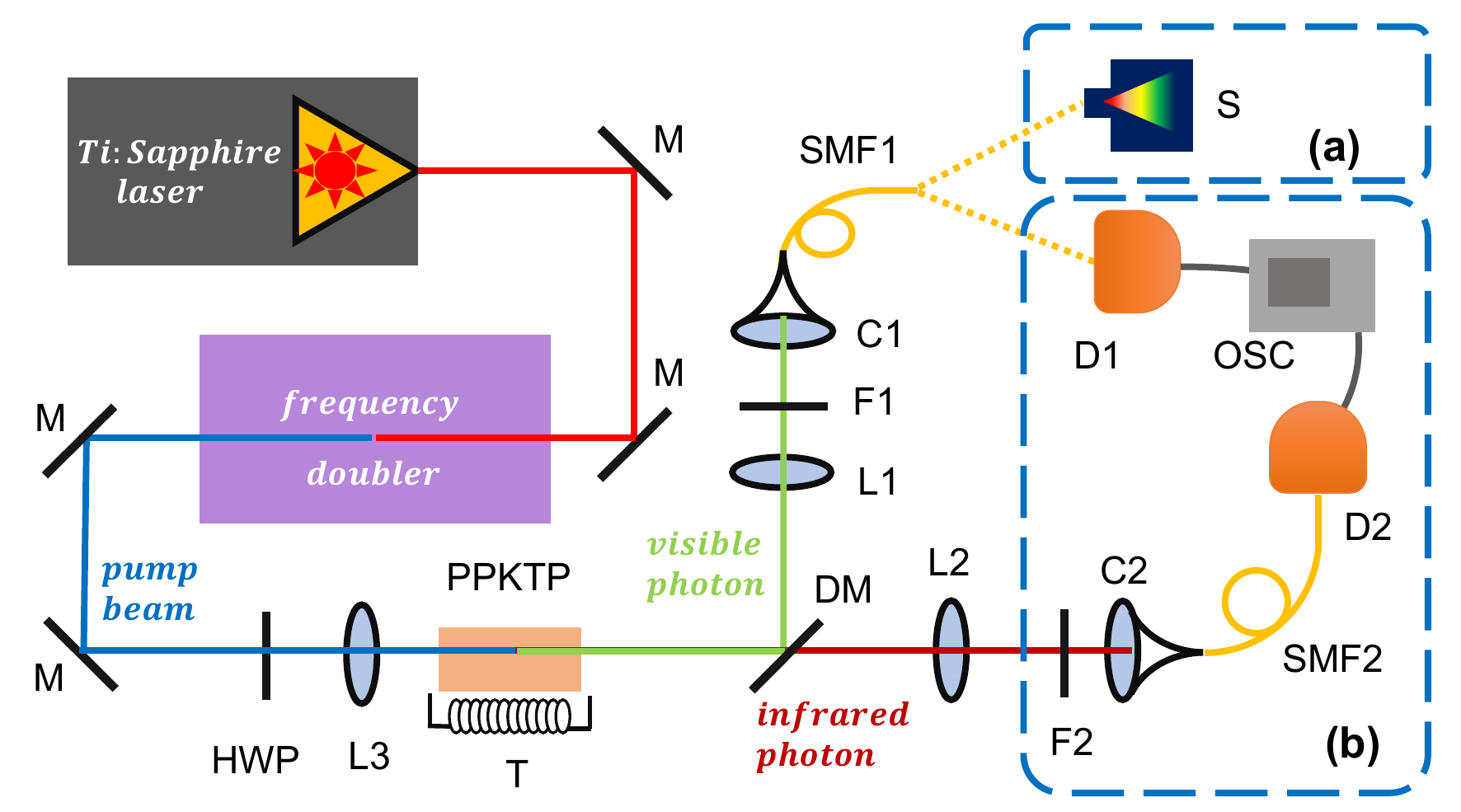}
		\caption[]{Experimental setup. Pulsed Ti:Sapphire laser, M -- mirror,  L3 -- lens (focal lengh f = 10 cm), PPKTP --  periodically poled potassium titanyl phosphate crystal, T --  temperature controller, DM -- dichroic mirror (Semrock 76-875 LP), L1, L2 -- plano-convex lens (f = $12$ cm, $15$ cm), F1 -- set of filters (three settings: (1) Chroma ET500 and Z532-rdc, (2) 2 pcs of AHF  442 LP, (3) 2pcs of AHF  442 LP and Thorlabs FESH0700), C1,C2 -- single mode fiber coupler (f = 0.8 cm, 1.51 cm), F2 -- long-pass filter (Semrock BLP01-1319R), SMF1, SMF2 -- single mode fiber ( two settings: (1) Thorlabs SMF460B and SMF1550, (2) SMF780 and SMF2000) Detection setups: a)  S -- spectrometer (Ocean Optics USB2000+), b) D1, D2 -- detectors, OSC -- oscilloscope. }
		\label{fig:setup}
	\end{figure}
	
\begin{figure}[ht]
		\centering
		\includegraphics[width=0.9\columnwidth]{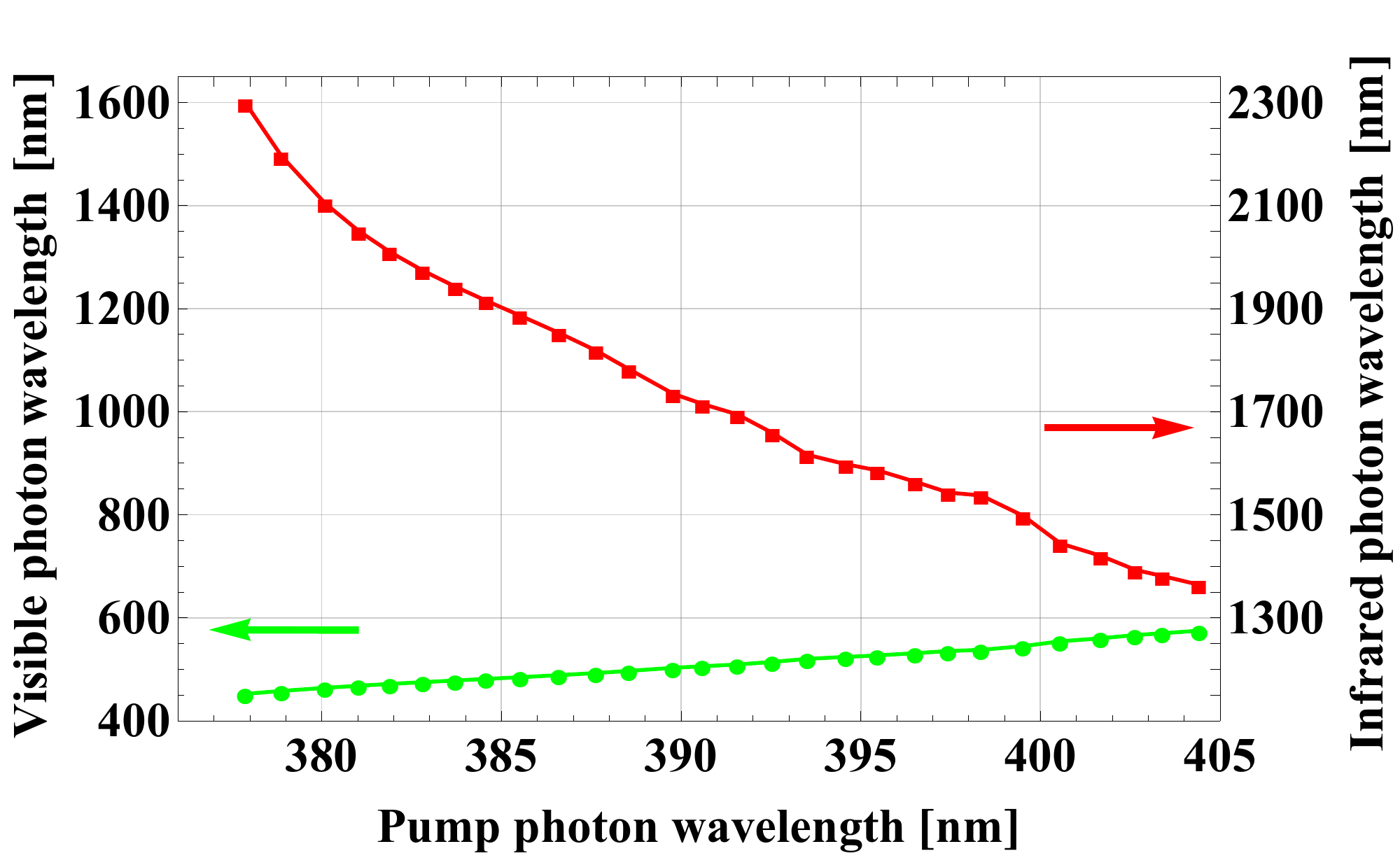}
		\label{fig:same_temp}
		\caption{ Generated photon wavelengths. Green dots show measured wavelengths of visible photon, red one are calculated from energy conservation relation.}
		\label{fig:tuningcurve}
	\end{figure}

	Finally, we tested our SSPDs with a correlated photon source based on the process of spontaneous parametric down conversion (SPDC), which we use to investigate their performance in correlation-type measurements\cite{Lutz2014,Misiaszek2018}. The experimental setup  is depicted in \figref{fig:setup}. A frequency doubled tunable Ti:Sapphire laser pumps the photon pair source based on a periodically poled potassium titanyl phosphate (PPKTP) crystal. The photons are coupled into single mode fibers. The spectra of the output photons can be modified by tuning the pump photon spectrum. The pump beam is tuned in the range of $377-404$ nm. It allows for generation of pairs of photons, where one photon is in $452-575$ nm spectral range and the other one in the $1363-2299$ range. The measured tuning curve is presented in \figref{fig:tuningcurve}. The pump and visible photon spectra are analyzed using a spectrometer. The spectrum of the infrared photon is calculated based on the energy conservation relation.

	\subsection{Results}
	
\begin{figure}[h]
		\centering
		\subfigure[]{
			\includegraphics[width=0.9\columnwidth]{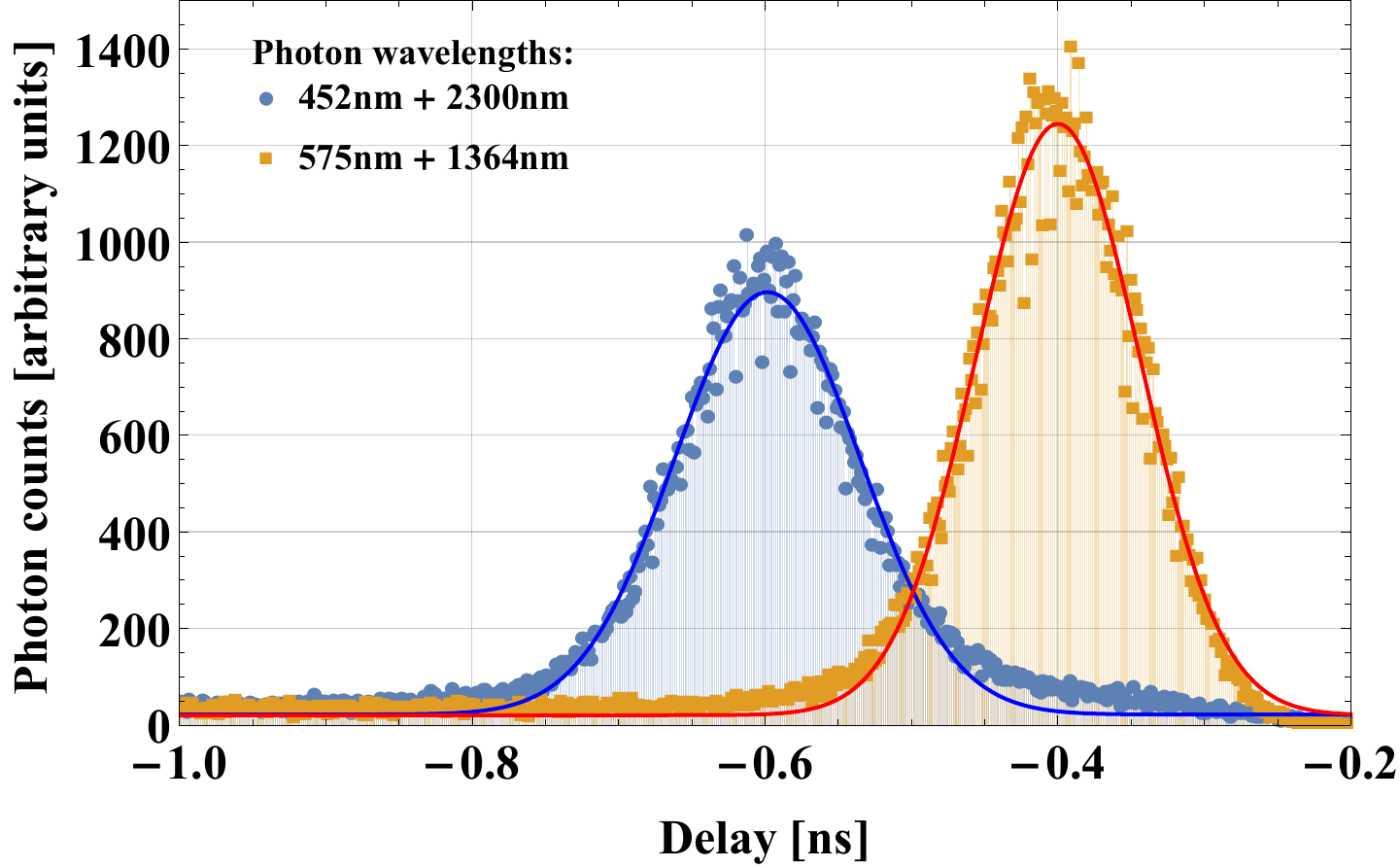}
		}
		\subfigure[]{
			\includegraphics[width=0.9\columnwidth]{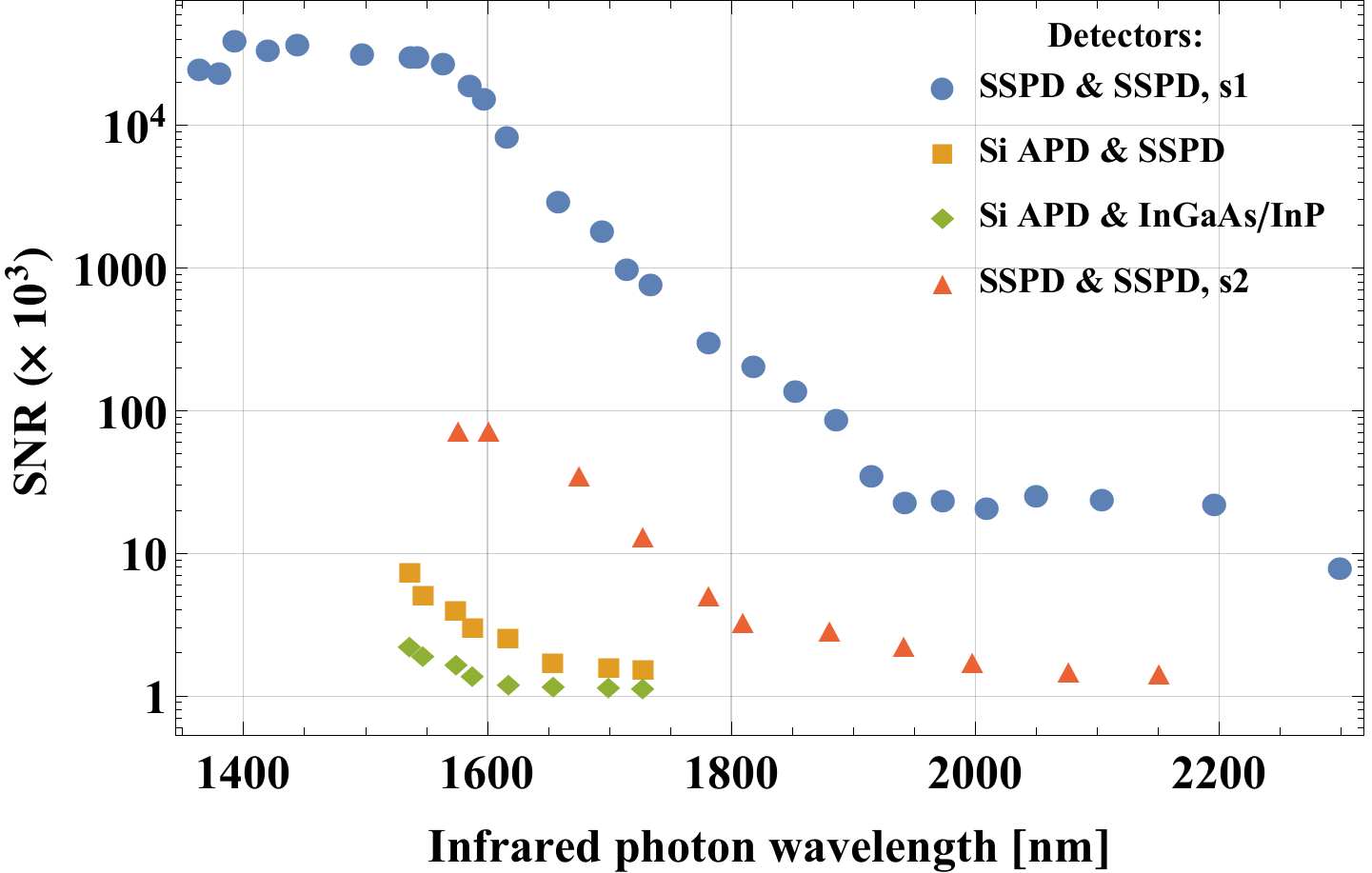}
		}
		\caption[]{(a) Arrival time coincidence histograms for two example settings. The measured setup timing jitter is approximately $86$ ps (FWHM). (b) SNR for four different settings as described in the main text. Plot markers are bigger than error bars, excluding SNR values close to 1.}
		\label{fig:SNR}
	\end{figure}

	The coincidence measurements are performed using three types of detectors: 1) SSPD: bias  current $21 \mu$A, 2) \mbox{InGaAs/InP}, detection efficiency $10\%$, gate width \mbox{$20$ ns}, deadtime $10\ \mu$s, and 3) silicon avalanche photodiode (Si APD).  The output signals from the detectors are then analyzed using an oscilloscope, \figref{fig:setup}(b). The timing histogram of the relative pulse arrival times are measured. The data set consists of $10^4$ samples. Figure \figref{fig:SNR}(a) shows examples of measurement results for different pairs of detectors: a) Si APD and InGaAs/InP, b) Si APD and SSPD and c) two SSPDs.

	The coincidence counting measurement  results, as seen in  \figref{fig:SNR} (a), allow us to estimate the signal to noise ratio (SNR) by fitting the Gaussian function \mbox{$b+a \exp[-4 \ln 2(t-t_0)^2/\sigma^2]$}  to the timing histograms. In our model  $b$ stands for the background noise, $a$ is the signal amplitude, $\sigma$ is the timing jitter and $t_0$ is the time where the pick maximum value occurs. Based on this definition  SNR $={a}/{b}{\sqrt{{\pi }/{16\log (2)}} \text{erf}(2 \sqrt{\log (2)})}\approx 0.52 {a}/{b} $. 
	
	The SNR results are gathered in panel (b) in \figref{fig:SNR}. We compare SSPDs with InGaAs/InP detectors in the exactly the same settings (F1 filter setting 1 and fibers setting 1 as described in caption of \figref{fig:setup} ). The results are depicted with green diamonds and orange squares. One can clearly see the limit around $1650$ nm beyond which the InGaAs/InP detector quantum efficiency drops to zero. The dark count rate of our Si APD was around $3$ kcps, which originated in an intrinsic dark counts of the detector itself and stray light. The  typical Si APDs' noise is around $100$ cps. Therefore, in principle, it is expected that SNR can increase in that case by one order of magnitude. Moreover, one can observe a plateau in the range $1900 -- 2200$ nm, which we attribute to the interplay of SDE for visible and infrared photons.
	
	Next, we use two SSPD detectors with different filters (F1 filter as denoted in setting 2 in \figref{fig:setup} ). The measurement results are marked using red triangles. This filter allows us to investigate the system performance in the broad range of wavelengths up to $2151$ nm. One can also see the improved SNR by approximately one order of magnitude. In the following step the SNR was improved by approximately 2 orders of magnitude by replacing the fibers (setting 2) and filters (setting 3). This allowed us to broaden the spectral range even further. It should be noted that in the range $1900-2300$ nm the SSPDs efficiency drop can be explained by small photon energies. 
	
	Recently, in the Ref.~\cite{Smirnov2018} we experimentally showed that the fabrication of SSPD detectors based on structurally disordered films \cite{Baturina2007, Gantmakher2010, Chockalingam2008, Chockalingam2009} allows for saturated dependence of the detection efficiency at lower bias current. This is based on the observation that the critical temperature and the electron diffusion coefficient follow the trend of the Ioffe–Regel parameter \cite{Vodolazov2017}, which is a measure of the degree of disorder.  Therefore this mechanism has the potential to increase the internal photon detection efficiency of an SSPD for the infrared photon wavelength range.


	\section{Conclusion}
	
	Summarizing, we have developed the principles of the realization of the single photon receivers above telecom wavelength range based on NbN SSPD with high values of system detection efficiency.  It was achieved by using more disordered NbN films. 
	The best values of the system detection efficiency are: $60$\% at $1700$ nm, $25$\% at $2000$ nm, $15$\% at $2300$ nm. 
	Within the experimental scenario involving the entangled photons we have compared our SSPDs with other detectors. The best performance was achieved when using an SSPD/SSPD pair, which yielded at least three orders of magnitude better SNR as compared to the others. In this case the achieved values of the SNR were $>20000$ up to $1600$ nm, $>1000$ in the wavelength range $1600-1700$ nm, $>100$ for $1700-1850$ nm, and $>8$ at around a $1850-2300$ nm.

	\section{Funding Information}
	
	\noindent \textit{PK} and \textit{MM} acknowledge support by: \\
	- National Laboratory of Atomic, Molecular and Optical Physics, Torun, Poland; \\
	- Foundation for Polish Science (FNP) (project First Team co-financed by the European Union under the European Regional Development Fund); \\
	- Ministy of Science and higher Education, Poland (MNiSW) (grant no.~6576/IA/SP/2016); \\
	- National Science Centre, Poland (NCN) (Sonata 12 grant no.~2016/23/D/ST2/02064). \\
	
	\noindent \textit{AD, YV, PM, KS, PZ }acknowledge support  by: \\
	- the Russian Science Foundation (RSF) Project No. 18-12-00364.

	
	\bibliography{FamoLab3.bib}
	

\end{document}